# Multi-peak vector soliton families in defocusing Kerr resonators

Pengxiang Wang[1,2], Carlos Mas-Arabí[3], Gian-Luca Oppo[4], Yiqing Xu[5], Miro Erkintalo[5], Stéphane Coen[5], Bertrand Kibler[6], Julien Fatome[6], and Gang Xu[1,2*]

[1]*School of Optical and Electronic Information, Huazhong University of Science and Technology, Wuhan, China*
[2]*Hubei optical fundamental research center, Wuhan, China*
[3]*Institut Universitari de Matematica Pura i Aplicada, Universitat Politècnica de València, València, Spain*
[4]*SUPA and Department of Physics, University of Strathclyde, Glasgow G4 0NG, Scotland, UK*
[5]*Department of Physics, The University of Auckland, Private Bag 92019, Auckland 1142, New Zealand*
[6]*Laboratoire Interdisciplinaire Carnot de Bourgogne, UMR6303 CNRS-UBFC, Dijon, France*

We report the existence of multi-peaked vector soliton families in normally dispersive passive Kerr resonators. Through cross-phase modulation between two orthogonal polarization components, each peak becomes tightly interlocked, enabling robust localization of the entire wave packet in defocusing cavities. Analysis using snakes-and-ladder diagrams demonstrates the diversity of these vector soliton families, which include dark-bright multi-peak solitons, flat-topped solitons, and modulation instability patterns, among others. Furthermore, stability analysis based on the coupled Lugiato-Lefever equations reveals that specific combinations of parameters can sustain stable vector cavity solitons, whose peak numbers can be continuously tuned by adding appropriate perturbations. These findings significantly expand the scope of soliton dynamics and optical frequency comb generation in pumped-dissipative systems, independent of dispersion conditions.

DOI:

*Introduction.* In optics and photonics, temporal cavity solitons (CS) are localized structures that manifest themselves as persisting pulses of light, and underpin numerous applications from ultra-short pulsed mode-locked lasers [1-3] to microresonator optical frequency combs (OFC) [4-6]. The refinement of OFC generation techniques has demonstrated unprecedented measurement accuracy and expanded new applications in a wide range of fields. This advancement now critically supports on technologies such as spectroscopy [7][8] precision atomic clocks [9] optical waveform synthesis [11] and promoted profound changes in quantum information manipulation [12].

In this framework, for the designing of high-performance optical resonators and the generation of high-quality OFCs, the ideal choices are short, bright and stable CSs [13-15] originating from the dual-balance of nonlinearity and dispersion, pumping and cavity losses [4][16]. Conventional wisdom dictates that stable bright cavity solitons (BCS) formation predominantly occurs under anomalous dispersion regimes [17,18], where the characteristic MI gain may essentially support the excitation of BCSs by efficient external writing or frequency-scanning techniques [19-21]. However, with normal dispersion, in most cases, the cavity field in normally dispersive resonators manifests themselves as homogeneous steady states (HSS) [22-24], Turing rolls [25-28] or spatial-temporal chaos [18][29][30], whose spectra are unsuitable for industry applications. From this point of view, guaranteeing the anomalous dispersion conditions is perquisite for high-quality OFC generation, which imposes strict constraints on both the center wavelength and resonator materials.

To this end, few theoretical and experimental investigations of dissipative solitonic pulses have been performed in recent years dealing with normal dispersion. At first, in the scalar case, stable dark pulses of different durations have been predicted and observed [31-33]. Next, the BCSs mediated by near-zero dispersion condition have been predicted and observed not only in fiber rings [34][35] but also in microresonators [36-38]. In order to extend the existence regime of stable solitonic pulses, cross-phase modulation (XPM) has been involved in with the aim of locking pulses [39][40]. Two mainstream approaches to CSs with XPM and normal dispersion are two- and three-coupled microcavities [41-43] and interlocked single-peak solitons supporting two circular polarization modes [15][40]. However, the full spectrum of these localized structures, their excitation mechanism, symmetry properties, and the existence and stability regimes remain unexplored.

In this work, we establish a unified framework for novel vector dissipative BCSs in normal-dispersion passive resonators. Unlike the stable multi-peak structures [44] or the localized patterns [45][46] in scalar defocusing cavity system, here, these XPM-induced localized structures, either bright or dark, can contain multiple peaks occupying certain ranges of the parameter space. By tuning the driving parameter of the cavity, these specific vector localized structures might evolve to be locked fronts (LF) [34][38][47] and polarization modulation instability (PMI) patterns [48-50]. More generally, we have revealed the essential and specific role of the XPM coefficient which may exert a significant influence on the threshold of field amplitude instability to the normally dispersive (or self-defocusing) systems.

*Model.* As shown in Fig.1(a), we consider a passive, coherently-driven Kerr ring resonator characterized by the superposition of two orthogonally polarized components $E_{1,2}$. Assuming high cavity finesse, the complex envelopes of the two modes obey the coupled Lugiato-Lefever Equations (LLE) [15][49][51-54]. In the dimensionless form, these equations read as follows:

$$\partial_t E_{1,2}(t,\tau) = \left[-(1+i\Delta) - i\partial^2\tau + i\left(|E_{1,2}|^2 + B|E_{2,1}|^2\right)\right]E_{1,2} + \sqrt{X/2}. \quad (1)$$

Here, $\tau$ and $t$ are the fast and slow time, describing the field evolution over a single-round trip and many round trips of the cavity, respectively. $\Delta$ is the cavity detuning, while $X$ is the total pump power and $B$ is the XPM coefficient.

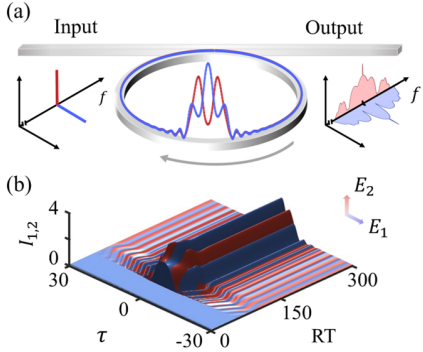

FIG. 1 (a) Schematic illustration of a normally dispersive resonator supporting bright vector solitons. The output frequency comb spectra correspond to a representative case, e.g. a 3–peak ($P_3$) BCS sustained in the fiber ring cavity. Red and blue curves (shadows) respectively represent the intensities (energies) in two orthogonal polarization modes. (b) The detailed stabilization evolution of $P_3$ BCS excited by intensity perturbations. On the 50$^{th}$ RT, a Gaussian perturbation is applied to the homogeneous steady states (HSS). After about 80 RTs, a stable $P_3$ BCS emerges in the resonator. The XPM coefficient $B = 1.85$ for all the cases in Figs. (1-3).

As shown in Fig. 1(b), the soliton excitation is realized by adding the intensity perturbation [55][56], leading to the subsequent PMI patterns [48] and the associated polarization spontaneous symmetry breaking (SSB) [15][47]. Eventually, they shrink to be stationary *N*-peak solitons (here $N = 3$). For more details of these novel vector solitons, we precisely calculate the intensity profiles by applying the Newton-Raphson method [57] and the continuation algorithms, thus providing an existence map shown in Fig. 2(a). Here, the characteristic boundary is defined by the left ($SN_1^l$) and right ($SN_1^r$) saddle-node lines. While the two solid black lines ($SN_h^l$ and $SN_h^r$) represent the saddle-node lines of the HSS [24] and the enclosed area corresponds to the bi-stable regime of the inclined resonances. Compared to the BCSs induced by the higher-order dispersion [34][35][58][59] or the Raman effect [60], these stable 3-peak vector BCSs could only be sustained at the bottom of the map in Fig. 2(a) (painted in red). In this regime, we have marked a typical spot for $\Delta = 5.5, X = 6$ by the green triangle and the corresponding intensity evolutions of the two polarization components are recorded in subpanels (i-ii). When $X$ is gradually increased and exceeds the threshold $X_{th} \approx 10.23$, the system passes the Hopf bifurcation boundary ($H_1$) and the BCSs enter into the oscillating regime (painted with orange), where the exemplary intensity evolutions (at $\Delta = 6.2, X = 8$, marked with blue triangles) are shown in subpanels (iii-iv).

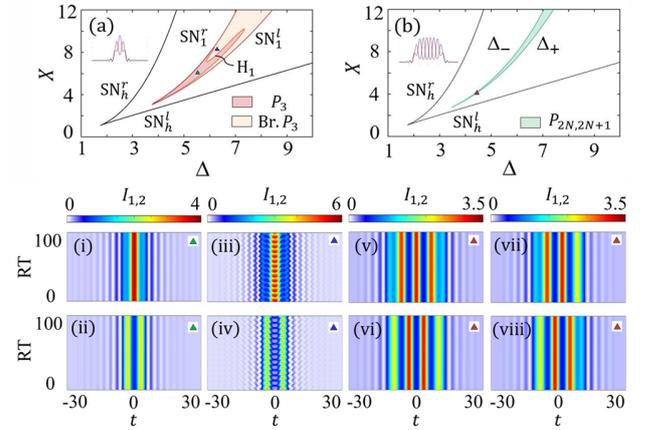

FIG. 2 (a). Existence map of stable and breathing $P_3$-BCS in th $\Delta$-$X$ space. The inset demonstrates the intensity profiles of the $P_3$ BCS. The subpanel shows the evolution of the $I_1$(i, iii, v, vii) and $I_2$(ii, iv, vi, viii). Green triangle (i-ii): stable $P_3$ BCSs with $\Delta = 5.5, X = 6$. Blue triangle (iii-iv): breathing $P_3$ BCSs with $\Delta = 6.2, X = 8$. (b) Map of the *pinning regime* (painted with light green) where stable vector soliton with multiple-peaks (>3) can coexist and the peak number of stable BCSs is proportion to the pulse width of the writing beam [see supplementary material]. Inset in (b) demonstrate the typical intensity profiles of an 8-peak localized structure. Red triangle (v-viii): stable $P_{2N+2}-$ or $P_{2N+1}-$BCSs ($N \geqslant 1$ is an arbitrary integer) with $\Delta = 4.35, X = 4$. The subpanels illustrate two exemplary evolution cases: (v-vi) $P_7 -$BCSs with temporal symmetry, (vii-viii) $P_6 -$ BCSs with time-reversed symmetry.

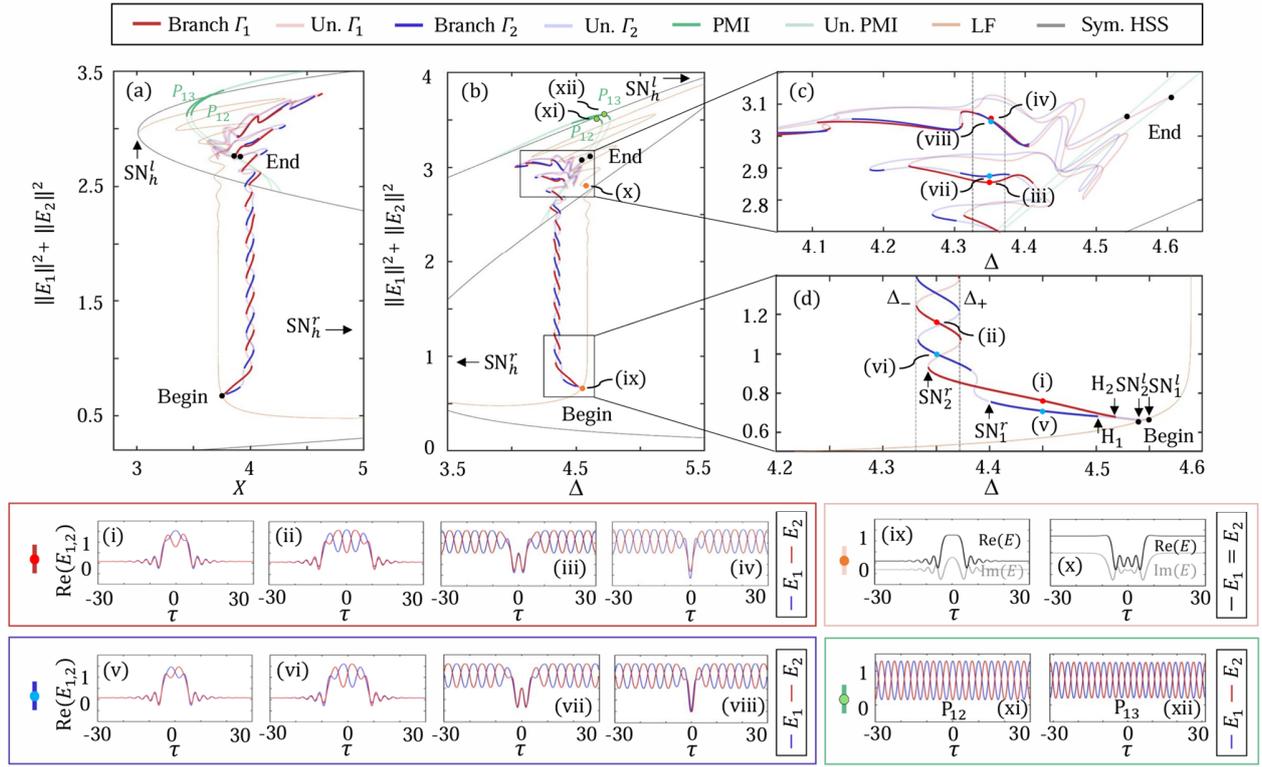

FIG.3 (a-d): Vector CSs snakes-and-ladders diagram for $B = 1.85$ and $\tau_{sp} = 60$. The averaged powers of the two polarization components are represented with red and blue curves, where the dark (pale) colors mean they are on the stable (unstable) states. (a): The $X$- bifurcation diagram for $\Delta = 4.35$ (b-d): The $\Delta$-bifurcation diagram for $X = 4$. Branches indicating normalized power for symmetric HSS (grey), LF (orange), temporal symmetry branch $\Gamma_1$ (red), time-reversed symmetry branch $\Gamma_2$ (blue) and PMI (green). All pale curves indicate the instabilities. (c-d): Zoom -in version of (b). Subpanels: example intensity profiles of $\Gamma_1$ bright solitons (i-ii), $\Gamma_1$ dark solitons (iii-iv), $\Gamma_2$ bright solitons (v-vi), $\Gamma_2$ dark solitons (vii-viii), LF (ix-x) and PMI patterns (xi-x).

More interestingly, one may obtain the distinct families of BCS with an arbitrary number of peaks ($> 3$), which may co-exist inside the specific pinning regimes $\Delta_- < \Delta < \Delta_+$ [painted in the clear green in Fig.2(b)]. More specifically, the stable BCSs could be categorized into odd- and even-peak classes, which reminds us of the precedent of the homoclinic snaking structure branches ($41/30 < \Delta < 2$) in the case of scalar LLE [61]. All of these structures are localized patterns embedded on HSS. Subpanels (v, vi) and (vii, viii) are evolutionary examples of odd peaks and even peaks, respectively. When $\Delta = 4.35$, $X = 4$, these structures can coexist and propagate in a stationary manner in the resonator. For larger pumps, although we find similar structures in the pinning regime through the continuation algorithm, their stability will eventually be lost due to the Hopf bifurcation [see supplementary materials]. In general, less-peak BCSs occupy wider existence range, beyond which, the fission of breathing BCSs occurs at their center part and the peak number is divided by 2. This cycle eventually gets ended when the peak number is sufficiently weak for the BCSs to stabilized themselves and supported inside the cavity.

In order to gain more physical insights of these distinct families of vector solitons, we analyze the complete vector soliton branches by plotting the cavity energy $[\|E_{1,2}\|^2 = \tau_{sp}^{-1} \int_{-\tau_{sp}/2}^{\tau_{sp}/2} |E_{1,2}|^2 d\tau]$ as a function of $X$ in Fig. 3(a) and $\Delta$ in Fig. 3(b − c), where $\tau_{sp}$ is the size of the fast time window. As shown in Fig. 3(a-b), the double-helix snakes and ladders diagram consist of two entangled soliton branches, which include the odd peaks ($\Gamma_1$ time symmetry branch) and even peaks ($\Gamma_2$ time-reversed symmetry branch) cases in the fast-time domain.

With the aim of clearly interpret the characteristics of these branches, we demonstrate the zoom in version maps of continuation curves, including the branch source which appears as short bright solitons in Fig. 3(c) and the branch end which is characterized by dipped dark solitons in Fig. 3(d). Both of these soliton branches are originated from the orange curve which is symmetric LFs $[E_1(\tau) = E_2(\tau)]$. In Fig. 3(a-b), the middle part of the two branches appears as a ladder-like spiral sequence, as the continuation moves upward one 'step' (more energy), reflecting the emergence of a pair of peaks on each side of the structure. Fortunately, the BCSs with any number of peaks could be stable for the parameter $\Delta = 4.35$ and $X = 4$, which is represented by a solid red or blue line. As the number of peaks increases, the localized structure will fill the whole ring cavity, and stable vector dark solitons are formed by locking each other with the oscillating tails. It is also worth mentioning that the variation of the fast time window size $\tau_{sp}$ may slightly alter the snaking ladder diagrams in Fig. 3(a-c). Notably, for the limited case of $\tau_{sp} \to +\infty$, the two starting points of the snaking branches in Fig. 3(a-c) asymptotically approach each other and eventually merge to be one. Moreover, for larger $\tau_{sp}$, these two curves may be more tightly spiraled, thus leading to more pattern periods of the endpoints [$P_{12}$ and $P_{13}$ for $\tau_{sp} \sim 60$, see Fig. 3(xi-xii)]. From this point of view, in a loose sense, the multi-peak BCSs could be regarded as *localized stable PMI patterns* without filling the entire angular space of the whole cavity. More specifically, it corresponds to homoclinic orbits in the phase space, which results from the interweaving the unstable symmetric LFs with pattern manifold.

In the bottom panels of Fig. 3, we recorded the typical temporal profiles of the BCSs. The labels (i-xii) and frame colors (blue, red, orange and green) correspond to different locations and curves of the snake-and-ladder diagrams in Fig. 3(a-d). Firstly, two vector CSs branches (bright and dark) are shown in subpanel (i-viii): one is temporal symmetric case (in red frame) and corresponds to profiles with local extreme value [ $\partial_\tau E_{1,2}(\tau)|_{\tau=0} = 0$ ] at the midpoint [see Fig. 3 subpanels (i-iv)]. Secondly, the other one is time-reversed symmetric case (in blue frame), corresponding to vector amplitude profiles with mirror-symmetry [$E_1(\tau) = E_2(-\tau)$] about the midpoint [see Fig.3 subpanels (v-viii)]. Moreover, both branches originate from the symmetric LFs (in orange frame) [see Fig.3 subpanels (ix)] and end to join the PMI (in green frame). In the end, in simulations, $\tau_{sp} = 60$, so the blue and red branches merge into the PMI of 12 cycles ($P_{12}$) and 13 cycles ($P_{13}$) respectively, as shown in Fig. 2(c) subpanels (xi-xii). In addition, the breathing multi-peak solitons are also discovered with high pump power. More continuation curves in this regime are provided in the supplementary material.

In all the analysis described above, the XPM coefficient $B$ in Eq. (1) has been fixed at 1.85. In the following, for the purpose of generality, we systematically investigate the crucial impact of this key parameter. As we have mentioned, compared with the scalar case, the XPM of the two polarization components may provide an additional PMI gain to the ring cavity, which is an essential and primary cause of cnoidal waves ('soliton crystals' or 'Turing rolls') [62][63] and the subsequent multi-peak vector solitons in the current work. In this case, the PMI gain spectrum could be determined by the eigenvalues of the perturbation matrix. Baring this in mind, we scanned the HSS curves at different $B$-values and determined that the gain effect under XPM influence is clearly pronounced even in the absence of SSB [64][65]. Therefore, at the boundary of the gain region (no SSB), the eigenvalues that indicate the gain spectrum could be simplified to be

$$\lambda_\pm(\Omega) = -1 + \sqrt{I^2 - q^2 \pm 2BI(I-q)}. \quad (2)$$

Here, $q(\Omega) = (2+B)I - \Delta + \Omega^2$ and $I = I_1 = I_2$. Due to the spatial reversibility of the LLE, the maximum points of the gain spectrum are obtained by solving the derivatives at $\lambda_\pm = 0$. These frequency components correspond to the most unstable component in the gain spectrum, and the expression can be interpreted as the phase-matching condition:

$$\begin{cases} d_\Omega \lambda_+: & 2(B+1)I - \Delta + \Omega_{1,2}^2 = 0 \\ d_\Omega \lambda_-: & 2I - \Delta + \Omega_{3,4}^2 = 0 \end{cases}. \quad (3)$$

These equations reveal the contribution of the XMP to the phase-shift balance, thus resulting to the boundary curves of the PMI regime shown in Fig. S1 in the supplementary material.

Benefiting from the well-defined PMI regime in the $X - \Delta$ plan, we may get access to generate stable BCSs with arbitrary number of peaks either by the detuning scanning or external writing (more details are in the supplementary material). As shown in Fig. 4(a), we take the typical cases for $P_5 -$ to $P_8 -$ BCSs as the initial condition and the subsequent continuation process on the $B$ coefficient reveal that the BCSs branches turn to be a series of isolas. Similar to the continuation curves on either the $X$ or the $\Delta$ coefficient reported in Fig. 3(a,b), here, these rings restricted inside the pinning regime (between $B_+$ and $B_-$) also contain the Hopf bifurcation and the SN bifurcation. In general, the solitons with more peaks are more likely to be unstable. For example, with the specific case for $\Delta = 4.35$ and $X = 4$ reported in Fig. 4(a), when the peak number is higher than 5, the Hopf bifurcation point starts appearing at the $B_+$ edge. Therefore, the BCSs turn to be breathing at this regime (shown by yellow curves). By further increasing the peak number, the breathing regime gradually extends and one may rationally anticipate that the stable BCSs will no longer exist when the peak number exceeds a certain threshold.

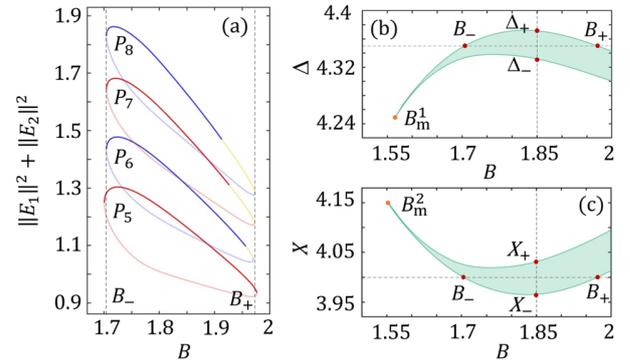

FIG. 4 Impact of the $B$ coefficient on the excitation and stability of the BCSs. (a). The $B$-continuation branch of $P_5$ - $P_8$ BCSs for $X = 4$ and $\Delta = 4.35$. Similar to those branches in Fig. 3(a-b), the red and blue curves refer to the even- and odd-peak cases. Dark (pale) color means the BCSs are stable (unstable). More specifically, the paled yellow part indicates that the BCSs are in breathing state. (b-c). The banana-shaped pinning regime of the $P_N -$BCSs for the fixed pump at $X = 4$ in the $B - \Delta$ plane and for the fixed cavity detuning at $\Delta = 4.35$ in the $B - X$ plane. The dashed horizontal lines and the red dots $B_+$ and $B_-$ refer to the specific scenario in (a). While the dashed vertical lines in (b) and (c), as well as the red dots correspond to the previous cases in Fig. 3(b) and Fig. 3(a), respectively.

More straightforwardly, to investigate the impact of the XPM, we plotted the pinning regime (painted in green) in function of the $B$ coefficient by fixing either the pump power [Fig. 4(b)] or the cavity detuning [Fig. 4(c)]. Very different to that in Fig. 2(a-b), here, the pining regime is deformed to be the banana shape, whose tips (marked with orange dots $B_m^1$ and $B_m^2$) refer to the limit for $B_+ = B_-$. These points can be regarded as the threshold of the $B$ coefficient for a given pump power or a given cavity detuning. Generally speaking, above these threshold points, by further increasing the $B$ coefficient, the pinning regime is expanding and the multi-peak solitons ($N > 3$) are more likely to be excited. However, we should also underline that, even inside this pinning regime, the soliton stability is still unguaranteed. As shown in Fig. 4(a), the stable regime is gradually shrinking by increasing the peak number.

In conclusion, we demonstrate the existence and the stability of the multi-peak vector solitons in passive Kerr resonators with normal dispersion. In this framework, the XPM of the two polarization components provides the net parametric gain, thus enabling generation of the BCSs with arbitrary peak number, which are reminiscent of the stable localized PMI patterns. The relevant continuation algorithm on the pump power, the cavity detuning, as well as the XPM coefficient reveal that these $N$-peak BCSs manifest themselves in the particular pinning regime. More specific analysis on the parametric gain and the soliton stability uncover their existence regimes and the parameter dependence, thus offering a generalized description of these novel soliton families.

For broader perspectives, this investigation lays the ground-work for operating Kerr bright soliton in simple-ring-cavity architecture with normal dispersion. It paves the way to overcome the long-standing obstacle to generate the OFC with short wavelength, e.g. in the visible or even the ultraviolet regimes, where the commonly-used materials for resonator waveguides are mostly with normal dispersion.

This work is partially supported by the National Natural Science Foundation of China (62275097) and the open project program of Hubei optical fundamental research center.

*gang_xu@hust.edu.cn

# Supplementary material to "Multi-peak vector soliton families in defocusing Kerr resonators"

Pengxiang Wang[1,2], Carlos Mas-Arabí[3], Gian-Luca Oppo[4], Yiqing Xu[5], Miro Erkintalo[5], Stéphane Coen[5], Bertrand Kibler[6], Julien Fatome[6], and Gang Xu[1,2*]

[1]*School of Optical and Electronic Information, Huazhong University of Science and Technology, Wuhan, China*
[2]*Hubei optical fundamental research center, Wuhan, China*
[3]*Institut Universitari de Matematica Pura i Aplicada, Universitat Politècnica de València, València, Spain*
[4]*SUPA and Department of Physics, University of Strathclyde, Glasgow G4 0NG, Scotland, UK*
[5]*Department of Physics, The University of Auckland, Private Bag 92019, Auckland 1142, New Zealand*
[6]*Laboratoire Interdisciplinaire Carnot de Bourgogne, UMR6303 CNRS-UBFC, Dijon, France*

In supplementary materials, we provide additional information on the mean-field model for the HSS solution, the analysis of the fast / slow time dynamics, the excitation techniques of the multi-peak solitons and the relevant stability analysis.

## I. Mean-field model and homogeneous steady state (HSS) solutions

We consider a Kerr nonlinear, passive, fiber ring resonator with two different orthogonal polarization modes which is coherently driven by a narrow-linewidth CW laser. In the slowly varying envelope approximations and mean-field limit, the complex amplitude $E_{1,2}$ propagating inside the cavity can be described by the coupled Lugiato-Lefever equations (LLE):

$$E_{1,2}(t,\tau) = [-(1+i\Delta) - i\partial^2\tau + i(I_1 + BI_2)]E_{1,2} + \sqrt{X/2}. \tag{S1}$$

In these coupled equations, $E_{1,2}$ represent the electric field complex amplitude in two orthogonal polarization components. When the intracavity intensity is below a certain threshold, the operation state of the resonator will be in the symmetry case ($E = E_1 = E_2$), and the system can be described by scalar LLE [1]. As the intracavity intensity increases, the symmetric solution loses the stability and the polarization spontaneous symmetry breaking (SSB) occurs ($E_1 \neq E_2$). The two polarization modes show different HSS solutions under the influence of cross-phase modulation (XPM), and a wealth of nonlinear dynamics and emerging waves are found in this situation, such as polarization domain walls (PDW), Polarization modulation instability (PMI) and multi-peak bright cavity solitons (BCS). In these coupled equations, the intracavity intensity $I_1 = |E_1|^2, I_2 = |E_2|^2$, $X$ is the pump or amplitude of the input field (plane-wave), and $\Delta$ is the frequency detuning to the closest cavity resonance. $t$ and $\tau$ is the 'slow time' and 'fast time'. In our study, the sign of dispersion is positive. The homogeneous $[\partial_\tau E_{1,2}(t,\tau) = 0]$ steady state $[\partial_t E_{1,2}(t,\tau) = 0]$ solutions of the coupled LLE satisfy a coupled cubic polynomial [3]

$$I_{1,2} = \frac{X/2}{1 + (I_{1,2} + BI_{2,1} - \Delta)^2}. \tag{S2}$$

In contrast to the familiar Airy equation of a nonlinear Fabry-Pérot resonator $I = aX/[1 + (bI - \Delta)^2]$, the solutions of Eq. (S2) complement the SSB path, which allows modes with different polarization directions to have mirror-symmetric power. Notably, when we force the two powers to be symmetric $[I_1 = I_2]$, Eq. (S2) degenerates into the Airy equation $[a = 1/2, b = 1 + B]$. The experimental results show that the system is bistable when the pump $X$ is in a certain range [2]. In fact, when $\Delta > \Delta_{th}$, the intracavity strength corresponds to three values, two of which are the stable solution and one is the saddle solution. Therefore, one can determine the range of the bistable state by looking for the position where the derivative $\partial_I X$ is equal to 0:

$$\partial_I X = 4(B+1)[(B+1)I - \Delta]I + [(B+1)I - \Delta]^2 + 1 = 0. \tag{S3}$$

The solution of Eq. (S3) corresponds to the two saddle-node points of HSS, which are depicted by the blue dashed curves ($SN_h^{l,r}$) in Fig. S1 and are given by

$$I_{1,2}^{SN} = \frac{2\Delta}{3(B+1)} \pm \frac{\sqrt{\Delta^2 - 3}}{3(B+1)} \tag{S4}$$

Next, we demonstrate the existence of a symmetry breaking solution. In order to more clearly analyze the relationship between the intensity of two polarization modes. By using two coupled cubic polynomial Eq. (S2) to eliminate variable $X$, we obtain

$$(I_1 - I_2)[(I_1 + I_2 - \Delta)^2 - (B-1)^2 I_1 I_2 + 1] = 0. \tag{S5}$$

The first term of the Eq. (S5) is a symmetric solution, i.e. $I_1 = I_2$, and the second term is the symmetric breaking case, indicating that the two intensities can satisfy an elliptic equation. Therefore, for the on and off positions of SSB (the intersection point of the symmetric solution and the broken solution of the elliptic symmetry), we can simply bring $I = I_1 = I_2$ into the elliptic equation [4] and get

$$I_\pm = \frac{2\Delta \pm \sqrt{4\Delta^2 - K(\Delta^2 + 1)}}{K} \tag{S6}$$

with $K = 4 - (B-1)^2$. In order to obtain the minimum power required by SSB in the parameter plane, Eq. (S6) is applied to Eq. (S2) and we simply obtain

$$X_\pm^{SSB} = 2I_\pm[1 + [(B+1)I_\pm - \Delta]^2]. \tag{S7}$$

As shown in Fig. S1, the SSB curve is represented by black dashed line. In this situation, if the stability and respiratory dynamics of SSB solutions need to be considered, time-dependent model should be taken into account, as in the references [5].

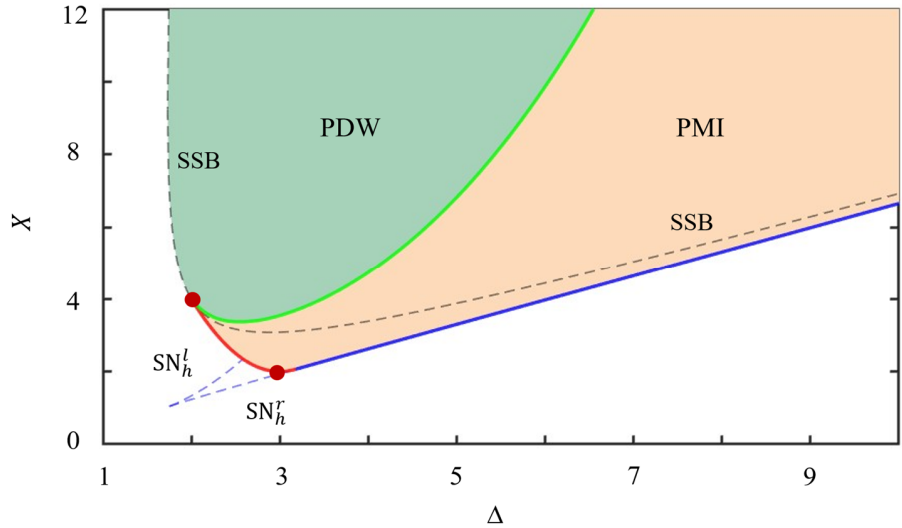

Fig.S1 **Characteristic regimes in the $\Delta - X$ plane**. The green and orange areas are PDW and PMI respectively, and the primary difference between the two is the gain effect. The PMI regime is enclosed by three curves: The red is the PMI gain threshold curve in for week detuning and pump. The blue curve is the upper saddle-node ($SN_h^r$) of HSS, while the green curve is the boundary between PMI and PDW, which means the change in the configuration of the spatial ($\tau$) feature root in the section II and the end of the temporal ($t$) gain effect in the section III. The black and blue dashed lines are SSB curves and saddle-node curves, respectively. The red dots represent the minimum pump thresholds ($X_{th}$) and detuning thresholds ($\Delta_{th}$) for the generation of the PMI structures.

## II. Numerical analysis for spatial ($\tau$) ODEs

In this section, we analyze the steady-state solutions, including continuous waves (CW), localized structures (LS), domain walls (DW), patterns, and their co-existing structures. We will also essentially study their possible coexistences, as well as their interconnections. To make it easy, it is reasonable to set the slow variable of the time envelope to be zero [$\partial_t E_{1,2} = 0$], and divide the complex amplitudes into real and imaginary parts to analyze separately, e.g. $E_{1,2} = E_{1,2}^r + iE_{1,2}^i$, so that the dynamics of the steady-state solutions could be described by a series of first-order ODEs as follows:

$$\frac{dE_1^r}{d\tau} = O \tag{S8}$$

$$\frac{dE_1^i}{d\tau} = P \tag{S9}$$

$$\frac{dE_2^r}{d\tau} = Q \tag{S10}$$

$$\frac{dE_2^i}{d\tau} = R \tag{S11}$$

$$\frac{dO}{d\tau} = -E_1^i + (|E_1|^2 + B|E_2|^2 - \Delta)E_1^r \tag{S12}$$

$$\frac{dP}{d\tau} = E_1^r + (|E_1|^2 + B|E_2|^2 - \Delta)E_1^i - \sqrt{X/2} \tag{S13}$$

$$\frac{dQ}{d\tau} = -E_2^i + (|E_2|^2 + B|E_1|^2 - \Delta)E_2^r \tag{S14}$$

$$\frac{dR}{d\tau} = E_2^r + (|E_2|^2 + B|E_1|^2 - \Delta)E_2^i - \sqrt{X/2} \tag{S15}$$

In these equations, $E_{1,2}^r = Re(E_{1,2})$ and $E_{1,2}^i = Im(E_{1,2})$. The linear stability of HSS in a spatial ($\tau$) space can be found in the eigen spectrum associated with the linearization of the dynamical system, which is obtained by solving the eigenvalue of the Jacobian matrix ($J_\tau$) around HSS. To simplify the analysis, we may assume that that each mode has a reference phase, which means the complex amplitudes are regarded as purely real [$E_{1,2} = E_{1,2}^*$]. With this approximation, the spatial eigenvalues follow the dependence [$|\lambda_\tau E_I - J_\tau| = 0$], which can be alternatively expressed as:

$$\begin{vmatrix} -1+T_1 & iI_1 & iK & iK \\ -iI_2 & -1-iT_1 & -iK & -iK \\ iK & iK & -1+T_2 & iI_2 \\ -iK & -iK & -iI_2 & -1-iT_2 \end{vmatrix} = 0 \tag{S16}$$

where $E_I$ is unitary matrix, $K = B\sqrt{I_1 I_2}$ and $T_{1,2} = 2I_{1,2} + BI_{2,1} - \Delta - \lambda_\tau^2$. An HSS solution corresponds to a group of 8 spatial eigenvalues, whose distribution in the complex plane may determine the spatial steady-state structures inside the cavity [5], including the CW, PDW, PMI and etc. As an example, the PDW and PMI described in Fig.S1 are two common steady-state solutions in this equation. These characteristics can be reflected in the distribution of the eigenvalues, and the boundary curve between PDW and PMI can be divided by different configurations, i.e. the solid green line in Fig. S1.

### III. Linear stability analysis in slow time ($t$) domain

In this section, we provide more details for the impact of the XPM on the phase-matching condition and the PMI gain spectrum. With this aim, we employ linear stability analysis to scan the Homogeneous Steady-State (HSS) curve for different values of the XPM coefficient $B$. Generally speaking, when the driving parameters are located inside the PMI regime in the Fig. S1, the HSS state inside the cavity would be destabilized by the exponentially-growing modulations, which will evolve to be periodic patterns or eventually the multi-peak solitons. Therefore, we perform the essential linear stability analysis of coupled LLE in in slow time ($t$) domain to demonstrate the evolution and propagation of perturbations on the steady-state solution. With this purpose, we apply added the perturbation term $E_{1,2} = E_{Hss} + u_{1,2}$ with the complex variable $u_{1,2} \propto \epsilon e^{\lambda t}$ and the stability of the HSS solution in slow time (for propagation) can be investigated with the gain spectrum represented by the following form [7]:

$$J_t = \begin{bmatrix} -1+S_1 & iI_1 & iK & iK \\ -iI_2 & -1-iS_1 & -iK & -iK \\ iK & iK & -1+S_2 & iI_2 \\ -iK & -iK & -iI_2 & -1-iS_2 \end{bmatrix} \tag{S17}$$

In this Jacobian matrix, $S_{1,2} = 2I_{1,2} + BI_{2,1} - \Delta + \Omega^2$, $\Omega$ refer to the normalized frequency component, and the corresponding parameter dependences are plotted in Fig. S2. In this figure, we demonstrate the bifurcation curves alongside their gain spectra, which is significantly affected by the XPM, even in the absence of the SSB. To make it clear, we put two examples with the cavity detuning $\Delta = 2.7$ (a-b) and $\Delta = 4.35$ (c-d). In both cases, by gradually increasing the pump power X, the PMI gain might appear before the occurrence of the SSB. As shown in Fig.S2. (a-b), the gain spectrum starts with intensity $I_-^{th} \approx 1.18$ and pump $X_{th} \approx 3.36$. For larger detuning (c-d), the starting position of PMI is replaced by the saddle-node curve ($SN_h^r$) of HSS, which proves that PMI is exactly occur on upper steady-state of HSS. Furthermore, we also plot the bifurcation curves of $B$ as a function of intracavity intensity and graph the gain spectrum with $B$, as shown in the Fig.S2. (e-f) respectively. Similarly, the gain spectrum begins with a symmetric solution, when $X = 4$ and $\Delta = 4.35$. In fact, at the bottom part of the PMI area (close to the red and blue curve in Fig. S1), the HSS for the two polarization components are symmetric, i.e. $I = I_1 = I_2$. In this case, for Eq. (S17), the eigen-polynomial $|\lambda_t E_I - J_t| = 0$ is degenerated biquadratic and the eigen values could be solved as follows:

$$\lambda_t(\Omega) = -1 \pm \sqrt{I^2 - q^2 \pm 2BI(I-q)}. \tag{S18}$$

Here, $q(\Omega) = (2+B)I - \Delta + \Omega^2$ and $I = I_1 = I_2$. Due to the spatial reversibility of the LLE, the maximum points of the gain spectrum are obtained by solving the derivatives at $\lambda_\pm = 0$. These frequency components correspond to the most unstable component in the gain spectrum, and the expression can be interpreted as the phase-matching condition:

$$\begin{cases} d_\Omega \lambda_+: \ 2(B+1)I - \Delta + \Omega_{1,2}^2 = 0 \\ d_\Omega \lambda_-: \ 2I - \Delta + \Omega_{3,4}^2 = 0 \end{cases}. \tag{S19}$$

These equations reveal that the phase shift balance is achieved with the joint contribution of the normal dispersion and the Kerr nonlinearities, which are related not only to the intracavity power and the cavity detuning, but also to the XPM. We should emphasize that in Eq. (S19), $\lambda_+$ and $\lambda_-$ can be interpreted as two co-existing gain ranges, which is quite different to the scalar LLE where only $\lambda_+$ is taken into account. For the gain of $\lambda_+$, the two specific frequency components $\Omega_{1,2}$ are the most unstable, and their maximum gain is $\lambda_+^{max} = (B+1)I - 1$. Therefore, the expression of the intracavity power threshold corresponds to the position where $\lambda_+^{max}$ is exactly zero. Note that if we take $B = 0$, then the threshold condition simplifies to $I_+^{th} = 1$ which represents the MI line in the scalar LLE. In the following, we mainly focus on the interpretation of $\lambda_-$, which is closely related the PMI gain. Similarly, by taking the condition $\lambda_-$ of Eq. (S19) into Eq. (S18), one may obtain $\lambda_-^{max} = (B-1)I - 1$, which is the maximum value at the $\Omega_{3,4}$. So, the two resulting thresholds from $\lambda_\pm^{max} = 0$ can be written as:

$$I_\pm^{th} = \frac{1}{B \pm 1}. \tag{S20}$$

When the cavity intensity exceeds this characteristic threshold $I_-^{th}$, the two polarization modes are affected by modulation, forming the PMI patterns that periodically alternate with each other. Combined with the mapping of conventional cubic equation of the bi-stable HSS, the boundary surface of PMI taking into account the B coefficient is represented as follows:

$$(B-1)X = 2\left[1 + \left(\Delta - \frac{B+1}{B-1}\right)^2\right]. \tag{S21}$$

This equation may provide essential information of the boundary curves for the bottom part of the PMI regime (shown in Fig. S1). More explicitly, as shown the red dots in Fig.S1, the pump threshold could be expressed as $X_{th} = 2/(B-1)$, which is given by the minimum value $[\Delta = (B+1)/(B-1)]$ of the quadratic function Eq. (S21). The detuning threshold $\Delta_{th}$ is located at the intersection point of the characteristic curve [red line, Eq. (S21)] and SSB boundary line [dashed black line, Eq. (S7)]. The top part of the PMI regime is enclosed with the boundary of PDWs (green line) [9].

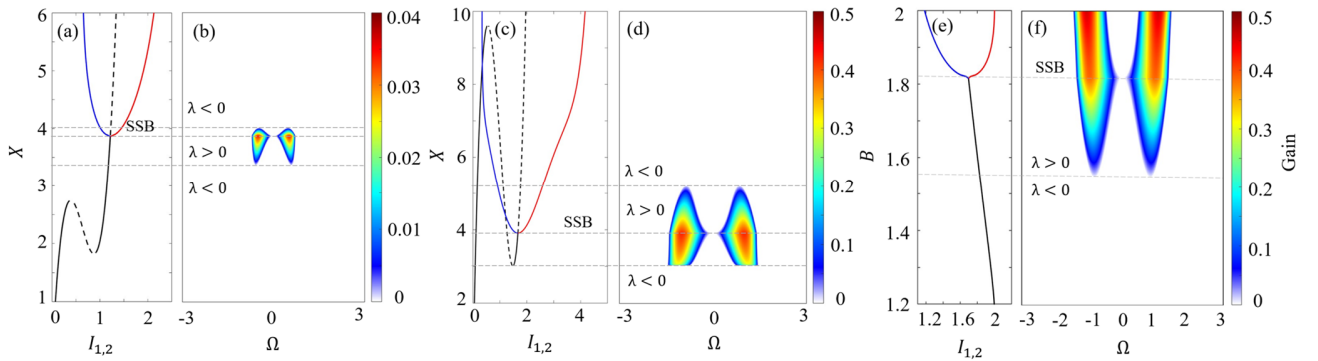

Fig.S2. **Branching curves of HSS solution and the corresponding gain spectrum.** (a-b) $\Delta = 2.7$, $B = 1.85$. The gain opening and closing thresholds point are on the red and green curves in Fig.S1. (c-d) $\Delta = 4.35$, $B = 1.85$. The gain opening and closing

thresholds correspond to the blue and green curves in Fig.S1. (e-f) Gain spectrum of resonator with different $B$ for $\Delta = 4.35$, $X = 4$.

## IV. The Origin of the Multi-Peak Soliton Branch

In this section, the origin of the multi-peak soliton branch is elucidated. In Fig. 4 of the main text, we presented the impact of the B parameter on the survival region of multi-peak solitons. To provide additional details, Fig. S3 depicts the multi-peak soliton branches for three distinct values of B: 1.85 (b), 1.6 (c), and 1.55 (d), in vivid analogues to the *different slices of the banana-shaped pinning regime*. Note that in Fig. S3(b-d), the vertical axis represents the total energy of the two polarisation modes, while the continuation parameter is the normalised pump $X$.

As shown in Fig. S3(b), we employed the parameters from Fig. 2 in the main text ($\Delta = 4.35$, $B = 1.85$). The red and blue curves exhibit a uniformly homoclinic snaking structure, indicating that such multi-peak solitons can be found between two specific values of the pumping power. As the B value decreases, the multi-peak soliton branch consistently maintains this homoclinic snaking structure, even as its survival interval $\delta_X = X_+ - X_-$ [the range where the curve folds in Fig. S3(b-d)] progressively narrows. Furthermore, $X_+$ and $X_-$ progressively approach to $X_M$ ($X_+ \approx X_- \approx X_M$) at $B = 1.55$. When $B$ falls below 1.55, the branch vanishes, meaning that the optical intensities in the two polarization directions become identical and the distinct multi-peak soliton solutions could no longer exist.

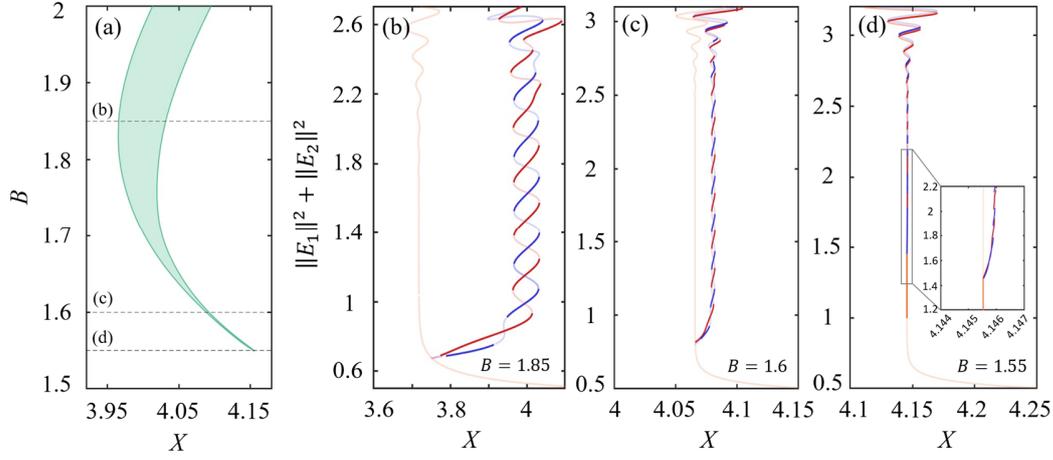

Fig.S3 **Homoclinic snaking branches with different parameter B.** (a) Existence regime of $P_N$ −BCSs in $(B, X)$ plane with $\Delta = 4.35$. (b-d) Continuation diagram of the soliton branch with $B = 1.85$, 1.6 and 1.55. The blue, red and orange curves represent $P_{2N}$-BCSs, $P_{2N+1}$-BCSs and LF respectively. The pale curves refer to the unstable solution.

## V. Excitation of stable multi-peak solitons

In this section, we demonstrate the excitation technics, including the detuning scan and the external-pulse writing, which enable the generation of stable vector multi-peak solitons.

(1) Detuning scan:

Our simulation illustrates the nonlinear dynamics of the system as the cavity detuning is varied according to the protocol shown in Fig. S4(a). This process consists of five stages (A-E). In stage A, the detuning gradually increases. It then reaches and remains fixed at a specific value in stage B. This procedure of scanning and holding detuning technique is a typical method for exciting bright cavity solitons (BCS) [8]. With this method, the intracavity intensity variations in two polarization modes are shown in Fig.S4 (b-c). The simulation starts from noise, which conveniently introduces asymmetry into the system. As the cavity detuning increases in stage A, the intensity difference between the two polarization modes becomes more pronounced. When the detuning linearly increases to 4.45 (within the existence regime of $P_2$- and $P_3$-BCSs), it is fixed, and the system enters stage B. This stage is characterized by the emergence of numerous kink structures. Around 50-60 roundtrips in Fig. S4(b-c), a structure similar to PMI, is observed. However, the system has just undergone a rapid change in detuning, which renders these modulations unstable. Stage C is where the PMI collapses. Due to the small intervals between kinks, adjacent peaks squeeze and compete with each other. Consequently, weaker peaks may decay back to the lower branch of the HSS, while stronger ones can eventually evolve into stable localized structures. The sequence from stage A to C constitutes the entire excitation process for $P_2$ or $P_3$ BCSs.

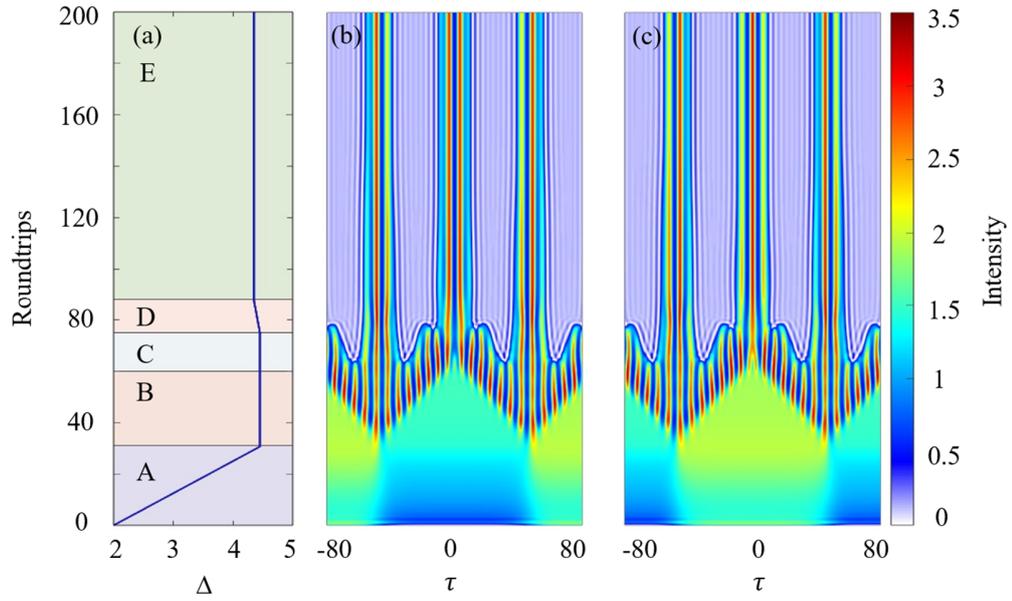

Fig.S4. **Multi-peak soliton excitation process by scanning frequency.** (a) The control tendency of frequency detuning. (b-c) Intracavity intensity $I_1$ and $I_2$ evolved in two polarization directions.

However, the excitation of multi-peak solitons requires two additional steps: stage D (a slight reduction of the cavity detuning) and stage E (relocking). According to the continuation curves in Figs. 3(a) and 3(b), the pinning regime lies on the side of the $P_2$-and $P_3$-BCS branches that corresponds to smaller detuning. Thus, the purpose of stage D in the sequence is to steer and confine the localized structures within the pinning regime; otherwise, they would all evolve into $P_2$- or $P_3$-BCSs. Finally, stage E consists of locking the detuning within the survival interval of the multi-peak solitons. As shown in Fig.S4(b-c), two $P_4 -$BCSs and one $P_5 -$BCS survive in the cavity.

(2) External-pulse writing:

Alternatively, stable vector BCS could also be deterministically written by external pulses by appropriately perturbing the homogeneous steady state (HSS) [9]. In passive Kerr resonators, the past experiments employ an incoherent "writing" scheme [10], in which an optical "addressing" pulse modulates the amplitude of the continuous-wave (CW) steady state via nonlinear cross-phase modulation (XPM). In this work, as shown in the Fig.S5 (b-c, e-f), on the 20th RT, two polarization modes are written into the excitation pulse, forming a domain wall (DW) structure under the action of nonlinearity. This DW continuously gets broadening over the next 100 RTs until the two polarization components differ significantly in intensity. After 150 RTs, the pulse width gradually stabilizes and tail structures emerge, which serve as a stable connection between the CW background and the periodic patterns.

By using this writing techniques, our numerical simulations show that $P_{2N}$-BCSs are the dominant outcome. This is due to the stringent formation condition for $P_{2N+1}$-BCS, which require symmetry about the center in the fast-time domain. If the difference in peak intensity between the two sides of $P_{2N+1}$-BCS exceeds a certain threshold, then it will eventually evolve to $P_{2N}$-BCS. To avoid this issue, we introduce asymmetries in the writing pulses for the two polarization modes. This scheme can not only determine the parity (even/odd) of the BCS peaks but also control the total number of peaks by adjusting the pulse width and the DW broadening period. More specifically, a slight intensity asymmetry between the two polarization components favors the formation of $P_{2N+1}$-BCS [Fig. S6(b)], whereas a temporal delay between them favors $P_{2N} -$BCS [shown in Fig. S6(d)]. Moreover, such intentionally introduced intensity asymmetry can effectively shorten the DW broadening period. Furthermore, an appropriate choice of the writing pulse width can control the number of peaks, as demonstrated in Fig. S6(a,c).

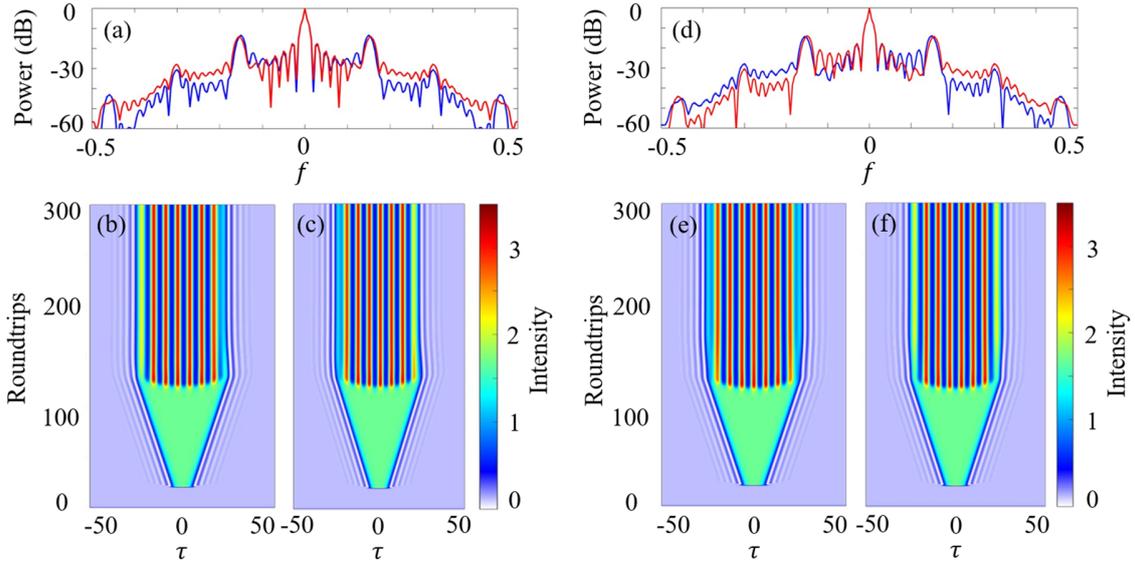

Fig.S5. **Multi-peak soliton writing by external pulses.** (a-c) Excitation case of $P_{2N}$-BCS. (a) Spectrum on the 300$^{th}$ RT. (b-c) Intensity $I_1$ and $I_2$ in two polarization modes of $P_{2N}$−BCS. (d-f) Excitation case of $P_{2N+1}$-BCS. (d) Spectrum on the 300$^{th}$ RT. (e-f) Intensity evolution of $I_1$ and $I_2$ of the two polarization components for $P_{2N+1}$-BCS.

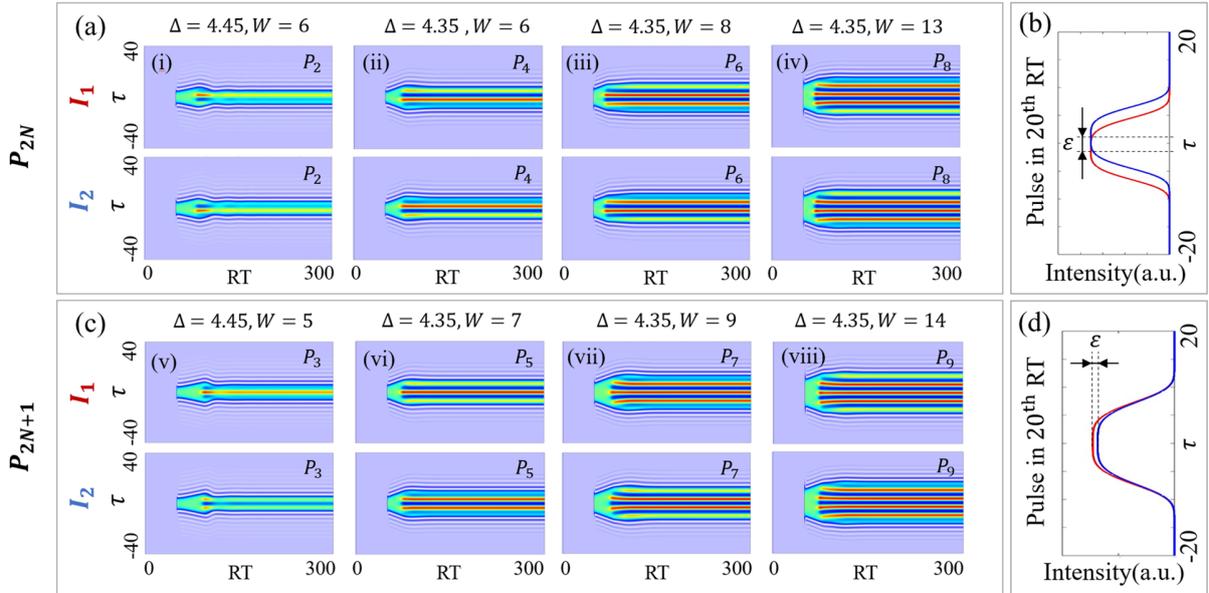

**Fig.S6. The excitation of BCSs with arbitrary peak number through Gaussian-shape pulse with different duration.** (a) $P_{2N}$-BCSs excitation evolution in 300 roundtrips. Subpanels (i) $P_2$: Write pulse width $W = 6$, detuning $\Delta = 4.45$. (ii) $P_4$: $W = 6$, $\Delta = 4.35$. (iii) $P_6$: $W = 8$, $\Delta = 4.35$. (iv) $P_8$: $W = 13$, $\Delta = 4.35$. (b) Excitation scheme for $P_{2N}$−BCSs: writing pulse with a delay in the fast time domain between the two polarization modes. (c) $P_{2N+1}$−BCSs excitation evolution in 300 roundtrips. Subpanels (v) $P_3$: Write pulse width $W = 5$, detuning $\Delta = 4.45$. (vi) $P_5$: $W = 7$, $\Delta = 4.35$. (vii) $P_7$: $W = 9$, $\Delta = 4.35$. (viii) $P_9$: $W = 14$, $\Delta = 4.35$. (d) The excitation scheme for $P_{2N+1}$-BCSs: writing pulse with a slight intensity difference in the two polarization components.

## VI. The oscillating dynamics of multi-peak vector solitons

Finally, we show the analysis addressing the identification and characterization of breathing BCS. With strong pumping, multi-peak BCSs undergo Hopf instability and period-doubling bifurcation. In this case, the multi-peak BCSs oscillate both in the temporal and spectral domains, as shown in the Fig.S7 (b-c, e-f). We select and plot a continuation curve for $X = 4.8$, where part of the soliton branch is replaced by an oscillating region (solid green line in the figure). Along this line, we characterize the eigenspectrum of a Hopf branch with a pair of imaginary eigenvalues $\lambda_t = \pm i\omega$ that leads to time oscillations. At this Hopf branch, solitons begin to oscillate in amplitude with frequency $\omega$.

We should also underline that vector solitons with different peak number could coexist inside the characteristic pinning regime. When increasing the intracavity power, the number of peaks increases as well. As shown in Fig. S7(a-d), the survival interval of stable BCSs narrows, allowing oscillating BCSs to dominate the pinning regime, while unstable BCSs (pale curves) also emerge. As the number of peaks and the total energy of solitons continued to increase, the breathing solitons were also insufficient to maintain their shape, and all the upper branches became unstable. Drawing from studies in the anomalous dispersion regime, third-order dispersion (TOD)/Raman effect—which typically causes drift instability—can suppress such oscillatory dynamics in BCSs [11][12]. Therefore, considering TOD in multi-peak solitons presents a promising approach for expanding their region of stable existence.

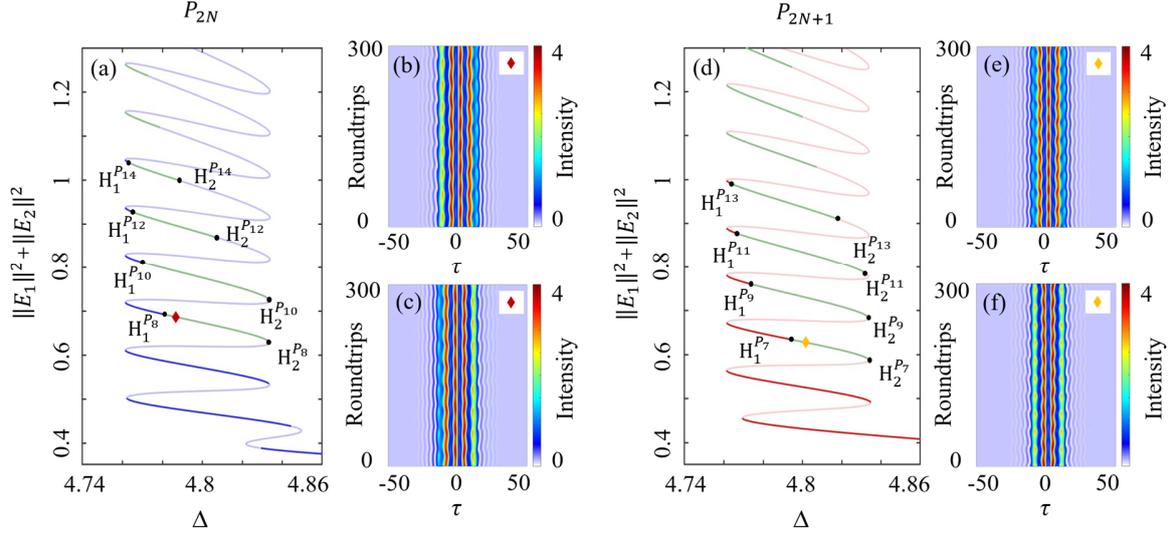

**Fig.S7. Oscillating multi-peak BCSs.** (a-c) $X = 4.8, \Delta = 4.785$ and $B = 1.85$. (a) $P_{2N}$-BCSs continuation curve takes energy as a function of detuning $\Delta$. The solid dark blue, pale blue and green curves are stable, unstable and breathing solitons respectively. $H_{1,2}^{P_{2N}}$ is the left and right Hopf bifurcation points of breathing $P_{2N}$-BCSs on the continuation curve. (b-c) Example of oscillating $P_8$ − BCSs with the parameters at the red diamond. (d-f) $X = 4.8, \Delta = 4.805$ and $B = 1.85$. (d) $P_{2N+1}$ −BCSs continuation curve takes energy as a function of detuning $\Delta$. The solid dark red, light red and green curves are stable, unstable and breathing solitons respectively. $H_{1,2}^{P_{2N+1}}$ is the left and right Hopf bifurcation points of breathing $P_{2N+1}$-BCSs on the continuation curve. (b-c) Example of oscillating $P_7$ BCSs with the parameters at the yellow diamond.

*gang_xu@hust.edu.cn